# General Formulation of Coulomb Explosion Dynamics of Highly Symmetric Charge Distributions

Omid Zandi[1,2*], and Renske M. van der Veen[1,2,3]

[1]*Department of Chemistry, University of Illinois at Urbana-Champaign, Urbana, IL 61801, USA*

[2]*Materials Research Laboratory, University of Illinois at Urbana-Champaign, Urbana, IL 61801, USA*

[3]*Department of Materials Science and Engineering, University of Illinois at Urbana-Champaign, Urbana, IL 61801, USA*

\* Corresponding author, email: omid.zandique@gmail.com

We present a theoretical approach to study the dynamics of spherical, cylindrical and ellipsoidal charge distributions under their self-Coulomb field and a stochastic force due to collisions and random motions of charged particles. The approach is based on finding the current density of the charge distribution from the charge-current continuity equation and determining the drift velocities of the particles. The latter can be used either to derive the Lagrangian of the system, or to write Newton's equation of motion with the Lorentz force. We develop a kinetic theory to include the stochastic force due to random motions of electrons into our model. To demonstrate the efficacy of our method, we apply it to various charge distributions and compare our results to *N*-body simulations. We show that our method reproduces the well-known emittance term in the envelope equation of uniform spherical and cylindrical charge distributions with correct coefficients. We use our model for the gravitational collapse of an ideal gas as well as the cyclotron dynamics of a cylindrical charge distribution in a uniform magnetic field and propose a method to measure the emittance of electron beams.

## I. INTRODUCTION

Coulomb explosion is a ubiquitous phenomenon resulting from the intense self-electric field of a charge distribution. The experimental and theoretical study of Coulomb explosion as a collective behavior of high-density charge distributions has been of interest in several different disciplines, including laser ablation [1][2][3], photoemission in electron sources and charged-particle optics [4][5][6][7][8], in streak cameras and microwave tubes [9][10][11], cold ion sources [12][13], plasma physics [14][15][16], chemical reactions [17], and space-charge imaging of molecules [18][19][20]. The characteristics of the charge distribution, i.e., its shape, uniformity, and initial velocity distribution, and the theoretical approximations necessary to describe its dynamics, depend on the specific application. Ellipsoidal charge distributions are of interest in the fields of high-brightness electron sources and optics for, e.g., X-ray free-electron facilities or ultrafast electron diffraction and microscopy setups [21][22][23][24][25]. Energy spread caused by Coulomb explosion of electron pulses can lead to energy spread of free-electron lasers [26]. The dynamics of photoemitted charged ellipsoids is often accompanied by a rapid transition from an initial oblate shape to a prolate shape at later times [27][28]. Uniformly charged ellipsoids can be spatially and temporally (re)compressed by linear electromagnetic fields to the limit that their emittance allows [29]. For more realistic non-uniform ellipsoidal charge distributions, the self-electric field will not be linear in space, which can result in shock waves in the distribution [30][32].

The dynamics of spherical charge distributions has attracted attention due to its importance in determining ion fragment distributions and their corresponding kinetic energies after a light-induced ionization of clusters [33][34][35][[36][37][38]. Often spherical charge distributions that are initially at rest are of interest. As they evolve (non-relativistically) over time, they maintain their uniformity because the self-Coulomb field remains linear in space. This approximate model simplifies the calculations and provides general insights in the physics of Coulomb explosion. While the dynamics of uniform spherical or cylindrical charge distributions are self-similar and can be described analytically [33][35], the Coulomb explosions of nonuniform ones are more complex [14][15][31][32] and, to our best knowledge, have not been studied for random motions of electrons and existence of an external magnetic field. When the granularity of the charge distribution is neglected, fluid models can be employed. The conservation of energy, for instance, has been used to find the distance of an ion separated from its original cluster by an intense laser pulse as a function of time [34][35]. Furthermore, the Vlasov-Poisson equations have been solved to model the dynamics of Coulomb explosion of nanotargets of different geometries (planar, spherical, and cylindrical) [38] or to find the kinetic energy distribution throughout the distribution [4].

Here, we present a new approach that facilitates the description of Coulomb explosion dynamics and can readily formulate the dynamics of any arbitrary (nonuniform) spherical or cylindrical charge distribution. The method takes advantage of the charge-current continuity equation to obtain the current density associated to a given charge distribution and is generated by the self-electric field of the distribution. The current density is a useful quantity by which the drift velocity field of the charges can be calculated. By the velocity distribution, the kinetic energy and hence, the Lagrangian of the system is found. The velocity distribution can also be used directly in Newton's equation of motion with the Lorentz force. We develop a kinetic theory to model the effect of electrons' random motions, i.e. a kinetic theory of emittance. Emittance is an important and usually limiting parameter in designing bright electron sources [8]. In this work, we refer to the collective effect of random motions of electrons (or mass particles in case of the gravity collapse) as a "stochastic process" that can be described by a force term in the equation of motion that governs the density evolution of the system, even when the system is far from equilibrium. Stochastic effects generate a thermodynamic pressure that can increase the rate of expansion during a Coulomb explosion. In a free explosion, the stochastic effects are usually insignificant. However, if the explosion is confined by, e.g., an external magnetic field, they become more pronounced due to the focusing effect of the magnetic field [8]. The first statistical analysis of electrons random motions is provided by Sacherer [39] and adopted by many others [8][40][41][42][43]. His analysis is based on the equations that the second moments of the phase space density satisfy in which the higher moments can be ignored if the root-mean-square (rms) emittance is constant or its time dependence is known a priori [39]. Here, we employ the kinetic theory of gases to include the effect of random motions in our model for arbitrary spherical or cylindrical charge distributions.

This paper is structured as follows. In Section II, we provide a general formulation of Coulomb explosion of spherical and cylindrical charge distributions with arbitrary densities. In Section III, we include the stochastic forces due to random motions of the electrons into our formulation and a kinetic theory of emittance is established. Several applications of this method are presented in Section IV where we first determine the dynamics of uniform ellipsoidal charge distributions in general, and those of uniform spheroid and spherical charge distributions in more detail due to their importance in many applications [8][27]. Additionally, we introduce the concept of self-conductivity that shows the response of a charge distribution to its own electric field. Then, we demonstrate how our method can provide an estimate for the statistics of nonuniform spherical charge distributions. A Gaussian charge distribution is examined for this purpose. Next, gravitational collapse of a spherical ideal gas as well as the dynamics of uniform cylindrical charge distributions in a magnetic field are presented as examples where the stochastic forces become significant because the distribution collapses. Many physical processes in astronomical scales are not accessible for controlled experiments. However, the similarity between the Coulomb dynamics and the gravitational dynamics has enabled the researchers to test them indirectly via controlled experiments with charged particle beams [44]. Here, we take advantage of this fact as well to formulate the collapse mechanism of an arbitrary spherical mass distribution and the effect of stochastic motions of mass particles. In Section V, we summarize our findings.

## II. THE GENERAL FORMULATION OF COULOMB EXPLOSION

For a charge distribution $\rho(\boldsymbol{r},t)$, we can find the corresponding current density $\boldsymbol{J}$ from the conservation of electric charge, i.e. the continuity equation $\nabla \cdot \boldsymbol{J} + \partial \rho / \partial t = 0$, then the drift velocity $\boldsymbol{v} = \boldsymbol{J}/\rho$, and write the non-relativistic equation of motion

$$m\frac{d\boldsymbol{v}}{dt} = -e(\boldsymbol{E} + \boldsymbol{v} \times \boldsymbol{B}) \qquad (1)$$

where $d\boldsymbol{v}/dt = \partial\boldsymbol{v}/\partial t + \boldsymbol{v}\cdot\nabla\boldsymbol{v}$ is the material derivative of $\boldsymbol{v}$, $\boldsymbol{E}$ is the self-electric field of the distribution satisfying the Maxwell equation $\nabla\cdot\boldsymbol{E} = \rho/\epsilon_0$, $\epsilon_0$ is the permittivity of free space, $\boldsymbol{B}$ is an external magnetic field and $m$ and $-e$ are the mass and charge of an electron, respectively.

Alternatively, we can form the nonrelativistic Lagrangian of the distribution as

$$\mathcal{L} = -\frac{1}{2}\frac{m}{e}\int |\boldsymbol{v}|^2\rho dV - \frac{\epsilon_0}{2}\int |\boldsymbol{E}|^2 dV + \int \rho\boldsymbol{v}\cdot\boldsymbol{A}\, dV, \qquad (2)$$

where the first and second spatial integrals are the kinetic and potential energies of the distribution, respectively, and **A** is the magnetic vector potential associated with the applied magnetic field and we write the Euler-Lagrange equation of motion.

By use of this approach for a charge distribution with spherical symmetry in the spherical coordinate $(r,\theta,\phi)$, with no external magnetic field, we find the nonlinear integrodifferential equation

$$\left[-\frac{\partial^2}{\partial t^2} + \frac{2}{\rho(r,t)}\frac{\partial\rho(r,t)}{\partial t}\frac{\partial}{\partial t} + \frac{e}{m\epsilon_0}\rho(r,t)\right]\int_0^r r'^2\rho(r',t)dr' = \left[\frac{2}{r^3\rho(r,t)} + \frac{1}{r^2\rho^2(r,t)}\frac{\partial\rho(r,t)}{\partial r}\right]\left(\frac{\partial}{\partial t}\int_0^r r'^2\rho(r',t)dr'\right)^2 \qquad (3)$$

governing the dynamics of the distribution. The derivation of Eq. (3) is provided in Appendix A. Solving this equation is beyond the scope of this work. The complexity of this equation indicates why simplified models are of great importance. Eq. (3) has a time-reversal symmetry, which means that if $\rho(r,t)$ is a solution to it, so is $\rho(r,-t)$. Also, the equation is invariant under sign reversal of the charge density.

Similarly, we show that the Coulomb explosion of a cylindrical charge distribution with an arbitrary profile in the cylindrical coordinate $(r,\phi,z)$, within a uniform constant axial magnetic field $B_0\hat{\boldsymbol{z}}$, is governed by the following set of three equations:

$$r\rho\left(\frac{\partial P}{\partial t}\right)^{-1}\frac{\partial v_\phi}{\partial t} + \frac{\partial v_\phi}{\partial r} + \frac{v_\phi}{r} = \frac{eB_0}{m} \qquad (4)$$

$$\left[-\frac{\partial^2}{\partial t^2} + \frac{2}{\rho}\frac{\partial\rho}{\partial t}\frac{\partial}{\partial t} + \frac{e\rho}{m\epsilon_0}\right]P = \left(v_\phi + \frac{eB_0}{m}r\right)v_\phi\rho + \frac{1}{r^2\rho^2}\frac{\partial(r\rho)}{\partial r}\left(\frac{\partial P}{\partial t}\right)^2 \qquad (5)$$

$$P(r,t) = \int_0^r \rho(r',t)r'dr' \qquad (6)$$

where $v_\phi$ is the azimuthal component of the drift velocity. The derivation of theses equations is provided in Appendix B. In derivations of Eq. (3)-(6), we did not have to distinguish multiple flow regions as is done in [4], since everything is encoded in the charge density function. Since we have used the current density of the distribution to find its dynamics, we may refer to it as the current density method.

### III.  EFFECT OF STOCHASTIC PROCESSES: KINETIC THEORY OF EMITTANCE

In this section we find stochastic forces due to random motions of electrons for spherical and cylindrical charge distributions discussed in the previous section. Suppose we can define the thermodynamic pressure $p$ that is the pressure due to random motions of electrons and not the mean field force $-e(\boldsymbol{E} + \boldsymbol{v} \times \boldsymbol{B})$, then we can write the equation of motion as [45]

$$m\frac{d\boldsymbol{v}}{dt} = -e\left(\boldsymbol{E} + \boldsymbol{v} \times \boldsymbol{B} + \frac{\nabla p}{\rho}\right) \tag{7}$$

However, to proceed, we need a state relation between this pressure and the electron density as we did for the mean field forces and the drift velocity. Again, we can assume that the explosion is an adiabatic process, and write the adiabatic state equation for ideal gases with three degrees of freedom [46]

$$p = p_0 \left(\frac{\rho}{\rho_0}\right)^{5/3} \tag{8}$$

where $p_0$ and $\rho_0$ are the pressure and the density at the initial time. Then the force due to the thermodynamic pressure will be

$$\boldsymbol{F}_{\text{th}} = -e\frac{\nabla p}{\rho} = -\frac{5}{3}\frac{ep_0}{\rho_0^{5/3}} \rho^{-1/3} \frac{\partial \rho}{\partial r} \hat{\boldsymbol{r}} \tag{9}$$

However, there are two problems with this approach: first, a system of charges undergoing coulomb explosion is far from equilibrium for which we may not be able to define thermodynamics states as in Eq. (8). In particular, the initial pressure may not even exist as we know that the emittance of a cold electron beam emerges and grows over time [8 ch.6]. Second, considering the uniform charge distributions, the force in Eq. (9) does not reproduce the well-known emittance term in the envelope equation. Indeed, for uniform distributions, this force is zero everywhere except at the edge of distribution, and hence is unrelated to the emittance which produces a linear force in space. Therefore, we need another theory that can, based on first principles, reproduce the emittance term. Our attempt in this section is to use the kinetic theory of gases with a fictitious boundary to model the random motions of electrons due to electron-electron collisions, which are elastic and conserve the momentum.

In later sections, we show that our theory here can reproduce the emittance term in the envelope equations with correct coefficients.

## A. Spherical Charge Distributions

In addition to the mean Coulomb force, a gas pressure exists due to random motions of electrons. To include the effect of random motions, we consider a fictitious sphere of an arbitrary radius $r$ with its center coinciding with the center of the spherical electron cloud. Electrons inside this sphere apply a pressure to electrons outside of the sphere due to their random motions as is depicted in Fig. 1 .

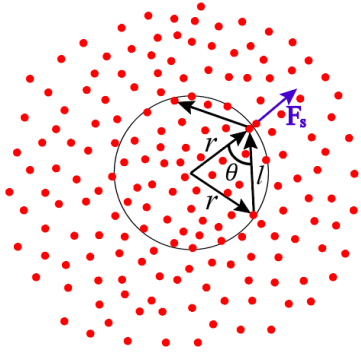

Fig. 1 Electrons within a spherical shell with a radius $r$ apply a force on the electrons outside of the shell due to their random motions. We consider a single electron that impinges on the inner side of the shell with an angle $\theta$ and reflects into the shell with the same angle with respect to the normal to the surface. Due to the conservation of linear momentum, the average force the electron applies on the shell is $2m\sigma_v \cos\theta/\delta t$, where $\delta t$ is the average time between two consecutive incidences on the shell and is $\delta t = l/\sigma_v$ and $\sigma_v$ is the spread of the electrons' velocities. From the geometry, it is obvious that $l = 2r\cos\theta$, so the force due to a single electron is $m\sigma_v^2/r$. The reflection is indeed caused by an electron-electron interaction which we modeled it by a fictitious boundary. However, as we are dealing with a fictitious boundary with an arbitrary radius, we should scale it with the fraction of total mass contributing to it at any radius.

Suppose an electron inside the sphere impinges on its interior surface and reflects with the same angle with respect to the normal to the surface. Some momentum is thus transferred to the surface area element. Because of the assumed symmetry, a plane that contains the trajectory of the electron before and after the collision, contains the center of mass of the distribution. Also, the force on the spherical shell lies on this plane. We therefore confine our derivations to this plane. It is straightforward to show that the force by the electrons inside the fictitious boundary on it is equal to $m\sigma_v^2/r$ where $\sigma_v$ is the velocity spread of electrons (see caption of Fig. 1 .) But this force should be scaled by the fraction of electrons that are contributing to it at radius $r$ i.e. $\xi = N\int_0^r r'\rho dr'/\int_0^\infty r'\rho dr'$. Therefore, the stochastic force for which we are looking becomes

$$\boldsymbol{F}_s = \xi m \frac{\sigma_v^2}{r}\hat{r}. \tag{10}$$

This scaling factor is zero at $r = 0$ as we expect this force to be zero at the center of mass.

Let's define $\eta^2 = N\sigma_v^2\sigma_s^2$ the phase space volume of the distribution on the plane on which we confined our derivation and $\sigma_s^2$ is the spatial spread of a single electron over a circle with radius $r$. The quantity $\pi\sigma_s^2$ is the spatial spread area of single electrons. It can be obtained by the expected area of the distribution at radius $r$ times the square of the spread of electrons around their average position, i.e. the squared coefficient of variation. Mathematically, it is given by

$$\pi\sigma_s^2 = \pi\langle r^2\rangle\frac{\langle r^2\rangle - \langle r\rangle^2}{\langle r\rangle^2}. \tag{11}$$

The bracket indicates the expectation value over the distribution. In other words, since an electron has no volume, if there is no uncertainty in their locations in the distribution, their spatial spread will be zero because in this case $\langle r^2\rangle = \langle r\rangle^2$. But because of the finite uncertainty in their locations, each point-like electron in Fig. 1 should be replaced by a (small) circle with radius $\sigma_s$.

It is straightforward to show that

$$\sigma_s^2 = \langle(r - r_D)^2\rangle, \tag{12}$$

where we define

$$r_D = \frac{\langle r^2\rangle}{\langle r\rangle}, \tag{13}$$

the dispersion radius of electrons after the statistical definition of index of dispersion that is the variance-to-mean ratio of a distribution. Eq. (12), in its expanded form is

$$\sigma_s^2 = -\frac{1}{Ne}\int (r - r_D)^2 \rho dV = -\frac{4\pi}{Ne}\int_0^\infty (r - r_D)^2 r^2 \rho dr. \tag{14}$$

Therefore, the velocity spread will be

$$\sigma_v^2 = \frac{\eta^2}{N\sigma_s^2} = -\frac{\eta^2 e}{4\pi\int_0^\infty (r - r_D)^2 r^2 \rho dr} \tag{15}$$

and the stochastic force in Eq. (10) becomes

$$\boldsymbol{F_s} = -m\frac{\hat{r}}{r}\frac{\int_0^r r'\rho dr'}{\int_0^\infty r'\rho dr'}\frac{\eta^2 Ne}{4\pi\int_0^\infty (r' - r_D)^2 r'^2 \rho dr'}. \tag{16}$$

Note that this force is always in the outward direction due to the sign of charge density. Also, although $N$ appears in Eq. (16), the stochastic force does not depend on it explicitly, as we expect for the emittance term. For a uniform charge density, $\boldsymbol{F_s}$ will be linear in $r$ as we again expect for the emittance term.

By including this force, Eq. (3) becomes

$$\left[-\frac{\partial^2}{\partial t^2} + \frac{2}{\rho}\frac{\partial \rho}{\partial t}\frac{\partial}{\partial t} + \frac{e}{m\epsilon_0}\rho\right]\int_0^r r'^2 \rho dr' = \left(\frac{2}{r^3\rho} + \frac{1}{r^2\rho^2}\frac{\partial \rho}{\partial r}\right)\left(\frac{\partial}{\partial t}\int_0^r r'^2 \rho dr'\right)^2 - \frac{\int_0^r r'\rho dr'}{\int_0^\infty r'\rho dr'}\frac{\eta^2 Ne\rho r}{4\pi \int_0^\infty (r'-r_D)^2 r'^2 \rho dr'}. \quad (17)$$

Eq. (14) represents the most general equation to describe the nonrelativistic dynamics of an arbitrary spherical charge distribution.

The phase space volume $\eta^2$ can be a function of time, as long as its evolution is considerably slower than the dynamics of the explosion. For an initially cold electron distribution, $\eta^2$ is zero initially, and while after the explosion, when the charge distribution is very dilute and we can ignore the electrons interactions, $\eta^2$ becomes a constant abiding by the Liouville's theorem.

## B. Cylindrical Charge Distributions

In general, for a freely expanding electron gas, the effect of the stochastic force is negligible in contrast to the Coulomb force. However, if the explosion takes place under the influence of a confining field, the internal pressure becomes more significant. For example, a cylindrically symmetric electron cloud that is radially confined by a uniform axial magnetic field experiences an enhanced internal pressure when the cloud periodically focuses (see Section VI. E and [28]), the size of which is influenced by the stochastic force.

The stochastic force can be added to Eq. (5) in a similar way as for spherical charge distributions. To do that, we will first find $\boldsymbol{F_s}$, which is the force due to random motions of electrons on a fictitious cylindrical shell with radius $r$ (and axis at $r = 0$). We ignore the effect of electrons motions in the $z$ direction because of symmetry and confine our derivations to a plane at $z =$ constant. Any electron that impinges on the cylinder will be reflected with the same angle as the incidence angle and applies a force along $\hat{r}$ on the cylinder wall. So, a single electron hitting the cylinder wall with an angle $\theta$ and the velocity $\sigma_v$ will apply an average force of $2m \sigma_v \cos\theta/\delta t$ on it, where $\delta t = 2r/\sigma_v \cos\theta$ is the average time between two successive collisions. This is the force that the electrons inside the cylinder apply to the electrons outside of the cylinder due to their random motions. Therefore, at distance $r$, $\boldsymbol{F_s}$ must be proportional to the portion of electrons contributing to this force, i.e. $N P(r,t)/P(\infty,t)$, which is the fraction of electrons within $r$. $P(r,t)$ is given by Eq. (6). Therefore, the force will be

$$\boldsymbol{F_s} = Nm\frac{P(r,t)}{P(\infty,t)}\frac{\sigma_v^2}{r}\hat{r}. \quad (18)$$

The functional form of $\sigma_v^2$ can be evaluated assuming that we know the phase space volume $\eta^2 = N\sigma_v^2\sigma_s^2$ for $\sigma_s^2$ being the spatial spread of single electrons within radius $r$. Similar to what we had in Eq. (12), we find

$$\sigma_s^2 = \frac{1}{P(\infty,t)} \int_0^\infty (r'-r_D)^2 r'\rho dr', \tag{19}$$

where $r_D = \langle r^2 \rangle / \langle r \rangle = \int_0^\infty r^2 \rho dV / \int_0^\infty r\rho dV$ is the electrons dispersion radius and hence

$$\sigma_v^2 = \frac{\eta^2}{\frac{1}{P(\infty,t)} \int_0^\infty (r'-r_D)^2 r'\rho dr'} \tag{20}$$

and the stochastic force becomes

$$\boldsymbol{F}_s = P(r,t) \frac{m\eta^2}{r \int_0^\infty (r'-r_D)^2 r'\rho dr'} \hat{r} \tag{21}$$

by use of which the radial component of the equation of motion becomes

$$\left[ -\frac{\partial^2}{\partial t^2} + \frac{2}{\rho} \frac{\partial \rho}{\partial t} \frac{\partial}{\partial t} + \frac{e\rho}{m\epsilon_0} \right] P = \left( v_\phi + \frac{eB_0}{m} r \right) v_\phi \rho + \frac{1}{r^2 \rho^2} \frac{\partial(r\rho)}{\partial r} \left( \frac{\partial P}{\partial t} \right)^2 + P(r,t) \frac{\eta^2 \rho}{\int_0^\infty (r'-r_D)^2 r'\rho dr'}. \tag{22}$$

In Section IV.E we will show that for a uniform cylindrical charge distribution, the stochastic term in Eq. (22) simplifies to the well-known emittance term of the distribution with a correct coefficient.

The forces we found in Eqs. (16) and (21) arise due to the random motions of the electrons, which originates from electron-electron collisions. Therefore, these forces are of stochastic nature. For the geometries we considered, this force is always along the Coulomb force and hence reinforces the expansion, meaning that collisions always increase the rate of expansion. The stochastic force is usually dominated by the Coulomb force when the expansion is unbounded. However, when there are other forces that can cancel the Coulomb force, like an external focusing force, the role of stochastic force becomes significant. In Section IV.E we will represent an example for this case.

## IV. APPLICATION OF THE CHARGE-DENSITY APPROACH AND THE KINETIC THEORY OF EMITTANCE

In this section, we demonstrate with simple examples how our approach in previous sections facilitates the description of Coulomb explosion dynamics.

### A. Ellipsoidal Uniform Charge Distributions

We consider an ellipsoidal charge distribution composed of $N$ electrons with the charge density

$$\rho(x,y,z,t) = -\frac{3Ne}{4\pi a(t)b(t)c(t)} H\left( 1 - \left( \left(\frac{x}{a(t)}\right)^2 + \left(\frac{y}{b(t)}\right)^2 + \left(\frac{z}{c(t)}\right)^2 \right) \right), \tag{23}$$

where $a$, $b$ and $c$ are the time-varying radii along axes $x$, $y$ and $z$, respectively, $e$ is the elementary charge, and $H(\cdot)$ is the Heaviside function

$$H(x) = \begin{cases} 0 & x < 0 \\ 1 & x \geq 0 \end{cases}. \tag{24}$$

The potential energy of the distribution is [47]

$$W = \frac{N^2 e^2}{10\epsilon_0} \frac{3}{4\pi abc} \int_0^\infty \frac{ds}{\sqrt{(1+s/a^2)(1+s/b^2)(1+s/c^2)}}, \tag{25}$$

and hence we can form the Lagrangian of the system as

$$\mathcal{L} = \frac{Nm}{10}\{\dot{a}^2 + \dot{b}^2 + \dot{c}^2\} - \frac{N^2 e^2}{10\epsilon_0} \frac{3}{4\pi abc} \int_0^\infty \frac{ds}{\sqrt{(1+s/a^2)(1+s/b^2)(1+s/c^2)}}, \tag{26}$$

For a uniform spheroid charge distribution with $a = b$, the Lagrangian simplifies to

$$\mathcal{L}_s = \frac{Nm}{10}\{2\dot{a}^2 + \dot{c}^2\} - \frac{N^2 e^2}{5\epsilon_0} \frac{3}{4\pi c} \frac{\cosh^{-1}(c/a)}{\sqrt{1-(a/c)^2}}, \tag{27}$$

which is valid for both oblate and prolate distributions. The corresponding Euler-Lagrange equations of motion are

$$\ddot{a} = \frac{Ne^2}{m\epsilon_0} \frac{1}{4\pi a^2} \Gamma_1\left(\frac{c}{a}\right) \text{ and } \ddot{c} = \frac{Ne^2}{m\epsilon_0} \frac{1}{4\pi a^2} \Gamma_2\left(\frac{c}{a}\right), \tag{28}$$

for

$$\Gamma_1(x) = \frac{3}{2} \frac{x - \frac{\cosh^{-1} x}{\sqrt{x^2-1}}}{x^2 - 1} \text{ and } \Gamma_2(x) = 3 \frac{-1 + x\frac{\cosh^{-1} x}{\sqrt{x^2-1}}}{x^2 - 1} \tag{29}$$

and $x = c/a$. The time-dependent radii of a uniform spheroid charge distribution composed of 10,000 electrons with initial radii $a(0) = 5\sqrt{5}$ μm and $c(0) = \sqrt{5}/10$ μm are calculated by numerically solving Eq. (28) and compared to $N$-body simulations in Fig. 2 (a). The stochastic force is not included in these calculations since, as the simulation results indicate, it has a negligible effect. The details of the $N$-body simulation are given in Appendix C. An excellent agreement between theory and simulation is obtained. An oblate-to-prolate shape transition occurs at $t \cong 95$ ps. This rapid shape transformation has recently been experimentally observed for a photoemitted electron cloud using an ultrafast electron microscope [28]. To explain the physics behind this transition, we consider the self-electric field of a uniformly charged spheroid

$$\mathbf{E} = -\frac{Ne}{4\pi\epsilon_0 a^2 c}\left(\frac{c}{a}\Gamma_1(c/a)r\hat{\mathbf{r}} + \Gamma_2(c/a)z\hat{\mathbf{z}}\right), \tag{30}$$

where $\hat{\mathbf{r}}$ and $\hat{\mathbf{z}}$ are unit vectors in the cylindrical coordinates. For $c/a \gg 1$, both functions $\Gamma_1$ and $\Gamma_2$, as plotted in Fig. 2 (b), become very small, with $\Gamma_1$ slightly larger than $\Gamma_2$. However, because of the factor $c/a$ in front of $\Gamma_1$ in Eq. (30), the electric field is significantly larger in the radial direction as expected. On the other hand, for $c/a \ll$ 1, the electric field will be mostly in the axial direction. The ratio of radial over axial components of the self-electric field in logarithmic scale is also plotted in Fig. 2 (b) vs. the aspect ratio. If a uniform spheroid charge density is initially very oblate, the electric field will be primarily along the small axis and hence the expansion will be significantly faster in that direction, eventually leading to a prolate shape.

The results of this method compare well to those of [27]. In our method, we take advantage of the charge-current continuity equation to find the drift velocity of electrons, which makes it possible to add electron sources. We also do not separate the oblate and prolate cases, as the Lagrangian in Eq. (26) (or Eq. (27)) models both cases.

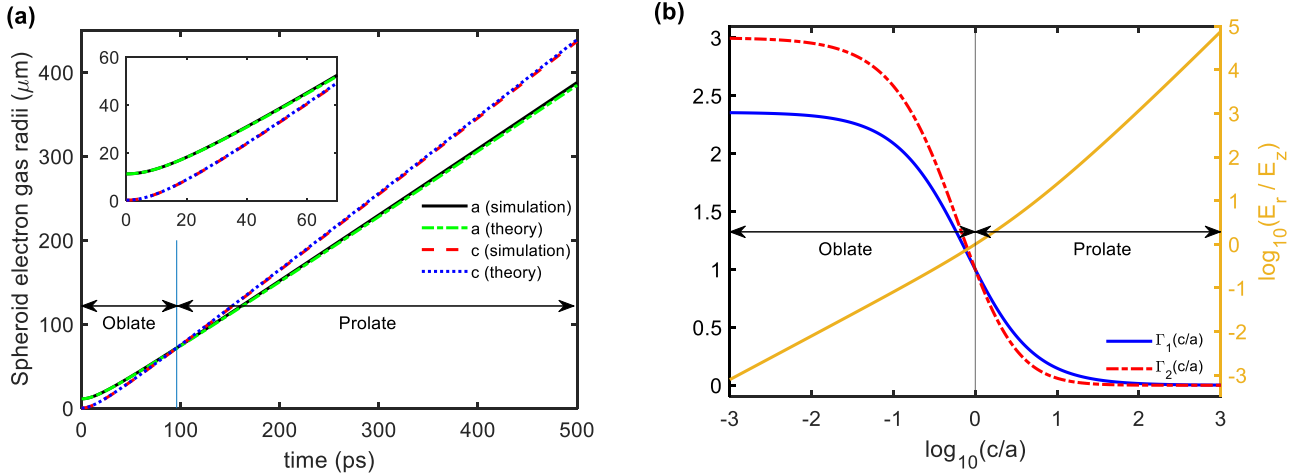

Fig. 2 (a) The long, $c(t)$, and short, $a(t)$, radii of a spheroid charge distribution expanding from rest under its self-Coulomb field determined by the current density method (theory) and *N*-body simulation. The distribution is composed of 10,000 electrons uniformly distributed within a spheroid with initial radii $a(0) = 5\sqrt{5}$ $\mu m$ and $c(0) = \sqrt{5}/10$ $\mu m$. (b) Functions $\Gamma_1(c/a)$ and $\Gamma_2(c/a)$ in Eq. (29) (left axis) vs. $c/a$ on a logarithmic scale. On the right axis is plotted the ratio of the radial component of the electric field over its axial component on a logarithmic scale. When the charge distribution is very oblate the electric field is primarily in the axial direction and vice versa.

## B. Uniform Spherical Charge Distribution

For a spherical charge distribution

$$\rho(r,t) = \frac{-3Ne}{4\pi R^3(t)} H\left(1 - \frac{r}{R(t)}\right) = \frac{-3Ne}{4\pi \left(5\sigma_r^2(t)\right)^{3/2}} H\left(1 - \frac{r}{\sqrt{5}\sigma_r(t)}\right), \tag{31}$$

where $R(t)$ and $\sigma_r(t) = R(t)/\sqrt{5}$ are the time-varying radius and standard deviation (std) of the electron cloud, respectively, Eq. (3) simplifies to

$$\frac{d^2 R}{dt^2} = \frac{Ne^2}{4\pi\epsilon_0 m R^2} \quad or \quad \frac{d^2 \sigma_r}{dt^2} = \frac{Ne^2}{4\pi\epsilon_0 m 5^{3/2} \sigma_r^2}. \tag{32}$$

In the literature, this type of equation is called an envelope equation [39]. Eq. (32) has an exact implicit solution. If $\dot{\sigma}_{r0} = (d\sigma_r/dt)_{t=0} = 0$, then

$$t = \sqrt{\frac{3}{2}}\frac{1}{\omega_0}\left\{ ln\left( \sqrt{\frac{\sigma_r(t)}{\sigma_{r0}} - 1} + \sqrt{\frac{\sigma_r(t)}{\sigma_{r0}}}\right) + \sqrt{\left(\frac{\sigma_r(t)}{\sigma_{r0}}\right)^2 - \frac{\sigma_r(t)}{\sigma_{r0}}} \right\}, \qquad (33)$$

where $\omega_0 = \sqrt{\frac{3Ne^2}{4\pi 5^{3/2}\sigma_{r0}^3 m\epsilon_0}}$ is the initial plasma frequency and $\sigma_{r0} = \sigma_r(0)$. This equation is identical to what is reported in the literature [4][35][31] and we repeat it here for later uses. The solution for $\dot{\sigma}_{r0} \neq 0$ is discussed in [48]. Fig. 3 (a) compares the time-varying spatial std of a uniformly charged sphere filled with 10,000 electrons obtained from Eq. (33) to that from an *N*-body simulation. The comparison is done for two initial stds of 2 μm and 5 μm. The details of the simulation are given in Appendix C.

When $\sigma_r(t) \gg \sigma_{r0}$ for $t \to \infty$, Eq. (33) can be approximated by

$$\sigma_r(t) \approx \sqrt{2/3}\,\omega_0\sigma_{r0}t. \qquad (34)$$

Quantity $\sqrt{2/3}\,\omega_0\sigma_{r0}$ has the unit of velocity and is the asymptotic rate of expansion. Notice that the product of the initial plasma frequency and the initial std gives a dependence of $\sqrt{N/\sigma_{R0}}$. Therefore, the time scale of the growth of $\sigma_r/\sigma_{r0}$ is eventually set by the initial plasma frequency. For instance, the spatial std of a uniformly charged sphere, with an initial radius of $2\sqrt{5}$ μm containing 10,000 electrons at rest, will reach the final expansion rate of $4.765 \times 10^5$ m/s, obtained from the simulation, while for the same charge distribution $\sqrt{2/3}\,\omega_0\sigma_{r0} = 4.759 \times 10^5$ m/s. The value of $\sqrt{2/3}\,\omega_0\sigma_{r0}$ as a function of the initial density for three different numbers of electrons initially at rest is plotted in Fig. 3 (b).

Using Eq. (34), the interior electric field of a uniform spherical charge distribution becomes

$$\boldsymbol{E}(\boldsymbol{r},t) = -\frac{Ne r\hat{r}}{4\pi\epsilon_0\left(5\sigma_r^2(t)\right)^{3/2}} \xrightarrow[t\to\infty]{} -\sqrt{\frac{\epsilon_0\pi}{2N}}\frac{\left(m5^{1/2}\sigma_{r0}\right)^{3/2}r\hat{r}}{e^2 t^3}. \qquad (35)$$

The asymptotic electric field is inversely proportional to the square root of the initial charge density, which is counterintuitive as it cannot be seen directly from Gauss's law. In Fig. 3 (c), the logarithm of $d|\boldsymbol{E}|/dr$ versus time is plotted for both the *N*-body simulations and the analytical solutions. The charge distribution with ~16 times higher initial density produces a smaller electric field at ~15 ps after time zero and at the same distance from the center of mass of the charge distribution.

The conductivity tensor $\bar{\bar{\kappa}}$ is defined by $\boldsymbol{J} = \bar{\bar{\kappa}}\boldsymbol{E}$. It essentially shows the collective response to an electric field. If we use the self-electric field for a uniform spherical charge distribution, and the current density it generates, it will be straightforward to show that

$$\kappa = 3\epsilon_0\frac{\dot{\sigma}_r(t)}{\sigma_r(t)} \xrightarrow[t\to\infty]{} \frac{3\epsilon_0}{t}, \qquad (36)$$

which is a scalar quantity, and we call it the self-conductivity. Fig. 3 (d) shows the self-conductivity for two different initial charge distributions as functions of time, obtained from the simulation. In the same graph, the asymptotic value of the self-conductivity in Eq. (36) is shown, which is independent of the initial charge density. This means that, once the self-conductivity converges to $3\epsilon_0/t$, the electrons have almost reached their final speed and the self-electric field becomes negligible.

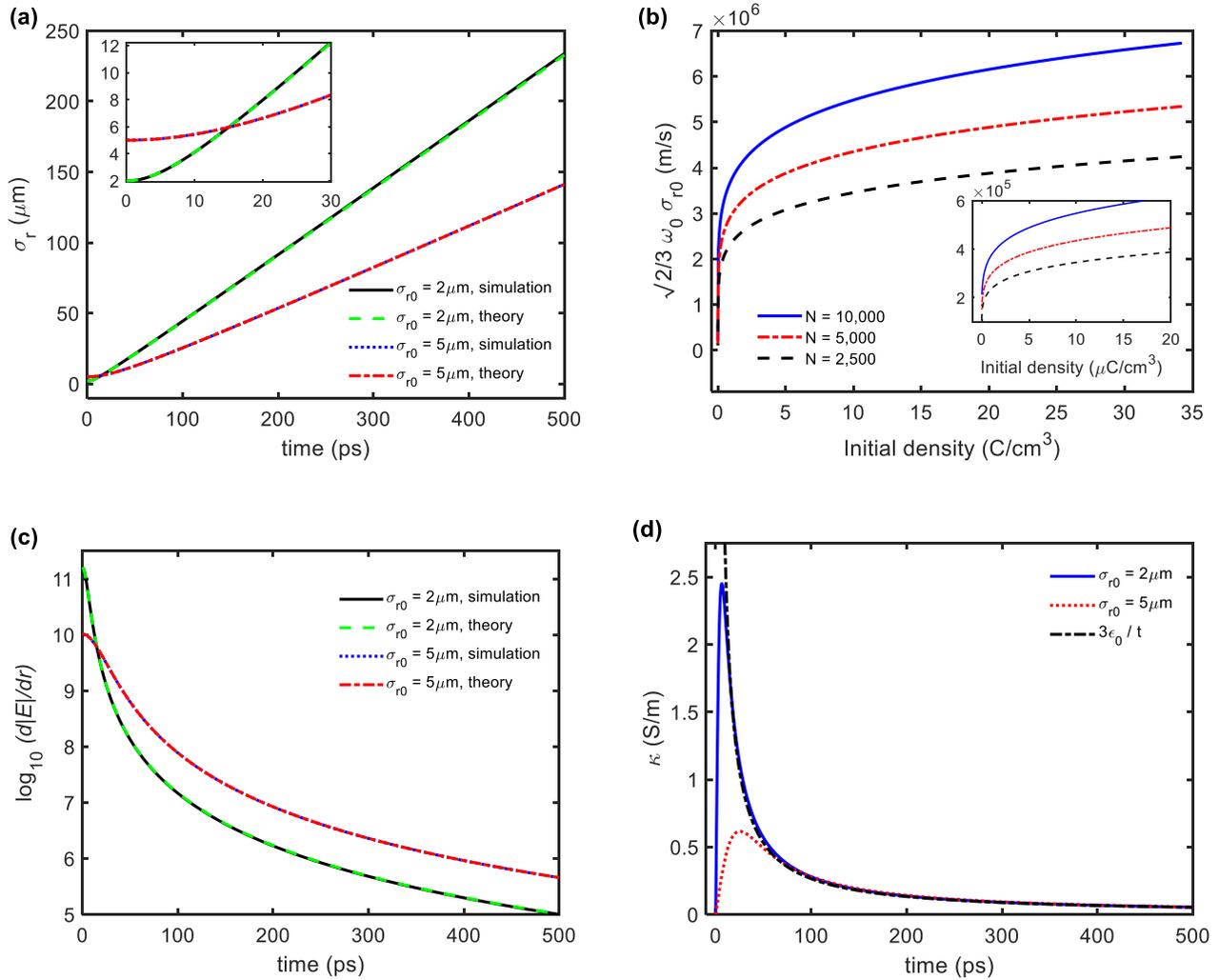

Fig. 3 Comparison of the simulated and the theoretical Coulomb explosion dynamics of a uniformly charged sphere containing 10,000 electrons starting from rest for two initial dimensions of 2 µm and 5 µm (std). (a) Spatial stds as a function of time obtained from *N*-body simulation and the analytical solution in Eq. (33). The inset shows a zoom into early time delays. (b) The final rate of expansion of the spatial std of the distribution as a function of initial charge density for three numbers of electrons. The final expansion velocities for lower initial densities are depicted in the inset. (c) $d|E|/dr$ on a logarithmic scale as a function of time, calculated from theory and *N*-body simulations. A higher initial charge density leads to a lower electric field at later times due to a faster expansion. Note that for uniform spherical charge distributions the electric field is linear in space inside the distribution (d) The self-conductivities of two freely expanding electron gases as a function of time. Regardless of the initial charge density, the self-conductivity always converges to $3\epsilon_0/t$ which is included in the figure for comparison.

Kinetic energy analysis is of great importance to obtain the energy distribution of charged particles, generated, for example, by light induced ionization of clusters [49]. Here, we show how the current-density method can be used for kinetic energy analysis. Using the drift velocity $\mathbf{v} = \mathbf{J}/\rho$ for spherical charge distributions, we can determine the highest kinetic energy that the electrons can obtain as

$$\beta(t) = \frac{1}{2}mv^2\left(t, r = \sqrt{5}\sigma_r(t)\right) = \frac{5}{2}m\dot{\sigma}_r^2(t). \tag{37}$$

Function $\beta(t)$ is plotted in Fig. 4 (a) for 10,000 electrons uniformly distributed in a sphere with an initial radius of $2\sqrt{5}\ \mu m$. From the electrons' kinetic energy and spatial distributions, it is straightforward to show that the number of electrons per kinetic energy is of the form

$$n(KE) = \alpha(t)\sqrt{KE}\ H\left(1 - \frac{KE}{\beta(t)}\right), \tag{38}$$

where $H(\cdot)$ is again the Heaviside function in Eq. (24), and $\alpha(t)$ is a proportionality function which can be readily obtained from the requirement that the integral of $n(KE)$ over all kinetic energies equates $N$, so

$$\alpha(t) = \frac{3}{2}N\beta^{-3/2}(t). \tag{39}$$

Therefore, the number of electrons with a kinetic energy in the range of $[KE, KE + \delta KE]$ is equal to

$$n(KE)\delta KE = \frac{3}{2}N\ \beta^{-3/2}(t)\ \sqrt{KE}\ H\left(1 - \frac{KE}{\beta(t)}\right)\delta KE. \tag{40}$$

This expression is equivalent to expressions given in [50]. Histograms showing the number of electrons per kinetic energy at four different points in time obtained from the simulation are plotted in Fig. 4 (b). Electrons start their dynamics from rest and the histogram bin width is $\delta KE = 50$ meV. In the same figure are also plotted theoretical curves calculated from Eq. (40). The deviation from the theoretical curve (square root distribution), specifically the peak near the highest kinetic energy, is due to granularity of the finite charge distribution which is thoroughly investigated in [51]. It is tempting to assign a temperature to the electron gas; however, the kinetic energy distribution is far from fulfilling Maxwell-Boltzmann statistics at all times. Since the average distance between an electron and its nearest neighbors increases in the order of $d = 2\sqrt{5}\sigma_r(t)/N^{1/3}$, the chance of collisions and hence thermalization decreases by time. The average and std of the kinetic energy are plotted in Fig. 4 (c), together with their theoretical values of

$$\overline{KE} = \frac{3}{5}\beta(t), \quad \text{and} \quad \sigma_{KE} = \sqrt{\frac{12}{175}}\beta(t), \tag{41}$$

respectively.

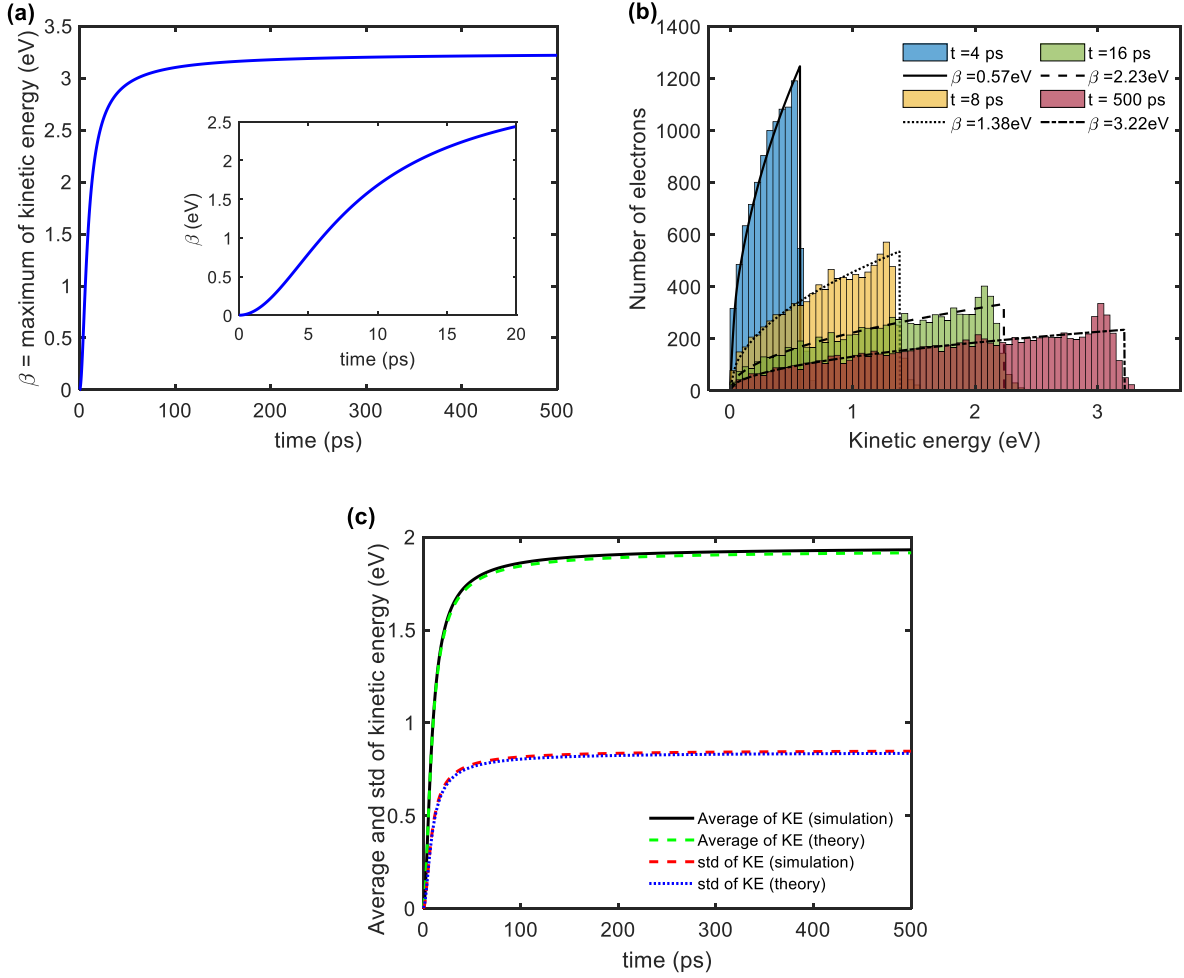

Fig. 4 (a) The maximum kinetic energy that electrons can acquire as a function of time determined by the mean-field theory in Eq. (37) for 10,000 electrons uniformly distributed in a sphere with an initial radius of $2\sqrt{5}$ μm. (b) Distribution of electrons vs. their kinetic energy at four different times. At each time, a curve obtained from Eq. (40) is also plotted. Deviations from the analytical curves correspond to deviations from mean-field theory and hence manifest random processes. The bin width for all four histograms is 50 meV. The peaks near the edge of the kinetic energy distribution are due to the granularity of the finite charge distribution and is discussed in [51] (c) Time-varying average and spread (sdt) of the electrons' kinetic energy. The change in the ratio of energy std over its average remains less than 1% after $t = 16$ ps.

In this section, we ignored stochastic effects in the equation of motion. Should we consider them, Eq. (17) for the uniform distribution in Eq. (31), becomes

$$\frac{d^2R}{dt^2} = \frac{Ne^2}{4\pi\epsilon_0 mR^2} + \frac{25\eta^2}{R^3}. \tag{42}$$

The term due to stochastic effects is proportional to $1/R^3$, as expected and the parameter $\eta$ is indeed the emittance of the charge distribution shown by $\epsilon$ in literature [8][39].

## C. Approximate Solutions for Statistics of Spherical Charge Distributions

In general, the integrodifferential equation in Eq. (3) is not solvable in a straightforward fashion, and hence approximate solutions can be employed to estimate the statistics of charge distributions over time. Except for uniform charge density, where both the electric field and the drift velocity are linear in space, no other charge distribution maintains its initial profile as it evolves by time because both its electric field and the drift velocity develop in a complex way. However, there exists a class of distributions for which the drift velocity and electric field are partially linear and hence we can use them to estimate their std as a function of time. In Appendix D, we show that all distributions of the form

$$\rho(r,t) = -\frac{3Ne}{4\pi\sigma_r^3(t)} f\left(\frac{r}{\sigma_r(t)}\right), \tag{43}$$

will result in a linear (in space) drift velocities, where $f(\cdot)$ is a unitless function whose volume integral over whole space is unity. Examples of this class of distributions are the uniform distribution, the Gaussian distribution, the raised cosine distribution, the triangle distribution, the Wigner semicircle distribution, and any other distribution whose std scales linearly with coordinates, among which only the uniform distribution maintains its profile as it evolves in time. Nevertheless, if the electric field is linear in the range $r \leq \sigma_r(t)$, we can estimate $\sigma_R(t)$ accurately abiding by Sacherer's conclusion that "the rms envelope equation depends only on the linear part of the forces" [39]. As an example, for a Gaussian charge distribution

$$\rho(r,t) = -Ne \frac{e^{-R^2/2\sigma_r^2(t)}}{(2\pi)^{3/2}\sigma_r^3(t)}, \tag{44}$$

it is straightforward to find the equation of motion for $\sigma_R(t)$

$$\dot{\sigma}_r^2(t) + \frac{Ne^2}{12\pi^{\frac{3}{2}}m\epsilon_0} \frac{1}{\sigma_r(t)} = \text{constant.} \tag{45}$$

which is accurate because in the range $r \leq \sigma_r(t)$ the electric field of the distribution is linear. However, we should emphasize that a Gaussian charge distribution does not maintain its profile as it evolves in time [31]. Therefore, the solution to Eq. (45) should not be inserted back into the Gaussian distribution to estimate how it evolves. Fig. 5 compares the spatial std of an initially Gaussian charge distribution to that obtained from an *N*-body simulation with 10,000 electrons, no initial velocity, and 5 μm initial spatial std. The Gaussian charge distribution expands slightly faster than the uniform charge distribution. The analytical model predicts the std of both distributions as a function of time well with a slightly higher accuracy for the uniform charge distribution.

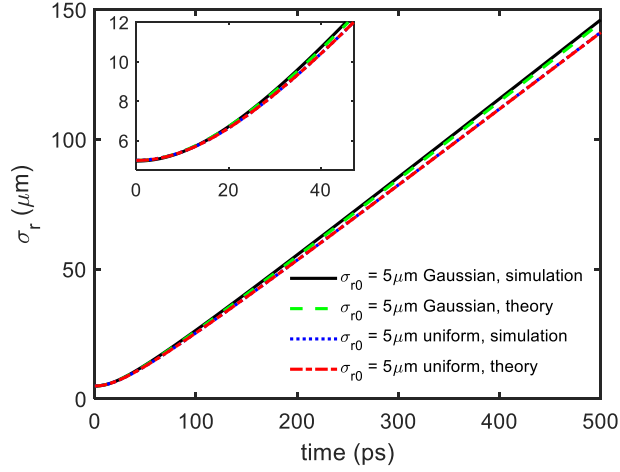

Fig. 5 Time-varying std of an initially Gaussian spherical charge distribution consisting of 10,000 electrons starting from rest, calculated using *N*-body simulation and theory. Dynamics of a uniformly charged sphere is included for comparison.

So far, we have ignored the stochastic force due to its negligible effect. In the next two applications of our method, we consider situations where this force must be considered.

## D. Gravitational Collapse of an Ideal Gas with Spherical Symmetry

Newton's gravity law has the same form as the Coulomb force. Therefore, our method can be employed to model the collapse of an ideal gas under its own gravity field. Unlike the Coulomb explosion, however, stochastic effects play a crucial role in determining the dynamics of the system and bringing it into equilibrium. Otherwise, the collapse would continue indefinitely. For a spherical mass distribution with a density $\rho_m(r,t)$, the collapse equation, using the Gauss's law for gravity field $\boldsymbol{E}_g$

$$\boldsymbol{E}_g = -4\pi G \frac{\hat{r}}{r^2} \int_0^r r'^2 \rho_m(r',t) dr', \qquad (46)$$

is

$$\left[ -\frac{\partial^2}{\partial t^2} + \frac{2}{\rho_m}\frac{\partial \rho_m}{\partial t}\frac{\partial}{\partial t} + 4\pi G \rho_m \right] \int_0^r r'^2 \rho_m dr' = \left( \frac{2}{r^3 \rho_m} + \frac{1}{r^2 \rho_m^2}\frac{\partial \rho_m}{\partial r} \right) \left( \frac{\partial}{\partial t} \int_0^r r'^2 \rho_m dr' \right)^2 + \frac{\int_0^r r' \rho_m dr'}{\int_0^\infty r' \rho_m dr'} \frac{\eta^2 M \rho_m r}{4\pi \int_0^\infty (r' - r_D)^2 r'^2 \rho_m dr'}. \quad (47)$$

where $G$ is the gravitational constant and $M$ the total mass of the distribution. The derivation of this equation follows the same procedure as for the Coulomb explosion of spherical electron gases represented in Appendix A, with the Newton's force of gravity and the stochastic force found in Section III. For a uniform spherical mass density

$$\rho_m = \frac{3M}{4\pi R^3} H\left(1 - \frac{r}{R}\right) \tag{48}$$

Eq. (47) simplifies to

$$\frac{d^2R}{dt^2} = -\frac{GM}{R^2} + \frac{25\eta^2}{R^3}. \tag{49}$$

For any nonzero value of $\eta$, the collapse will not continue indefinitely and reaches an equilibrium at which $\ddot{R} = 0$. The equilibrium radius will be $R_{eq} = 25\eta^2/GM$. In the collapse process, the radius of the uniform gas may oscillate around the equilibrium radius. Unlike the free Coulomb explosion, the stochastic (thermodynamic) force, which leads to the last term in Eq. (49), is of great importance to determine the dynamics of the collapse, since its sign is opposite to that of the gravity force term. Like the case of electrical charge distribution, here, we can define the self-conductivity as a tensor relating the gravity field to the mass current density $\boldsymbol{J}_m = \bar{\bar{\kappa}}_m \boldsymbol{E}_g$. For the uniform spherical ideal gas $\kappa_m = -\frac{3}{4\pi G}\frac{\dot{R}}{R}$. If the dynamics starts from rest, then $\dot{R}\big|_{t=0} = 0$. Also, at the equilibrium $\dot{R}\big|_{t\to\infty} = 0$; therefore, $\kappa_m$ will have at least one extremum in between, the time of which depends on the initial mass density and $\eta^2$. In general, the conservation of phase space volume is less strict for gravitational collapse because the collapsing mass consists of material elements that, unlike electrons, have internal structures/energy levels and due to friction and other processes part of their kinetic energies are transferred into internal energies. So far, we have assumed that for an ideal gas, the gravitational collapse of a dilute uniform spherical ideal gas remains self-similar. However, when an ideal gas with three degrees of freedom equilibrizes under its own gravity force, a thermodynamic force, similar to what we discussed in Eq. (9), arises, that is

$$\boldsymbol{F}_{\text{th}} = -M\frac{\nabla p}{\rho_m} = -\frac{5}{3}\frac{Mp_0}{\rho_{m0}^{5/3}}\rho^{-1/3}\frac{\partial\rho_m}{\partial r}\hat{\boldsymbol{r}} \tag{50}$$

where $p$ is the thermodynamic pressure, and $p_0$ and $\rho_{m0}$ are respectively the pressure and the density at some arbitrary time after the equilibration. For uniform densities, this force will not be linear in space which indicates that uniform distributions do not remain self-similar. Nevertheless, for uniform distributions, the thermodynamic force in Eq. (50) is nonzero only at the edge of the distribution, and for dilute ideal gases we may ignore it.

The gravitational collapse and the effect of granularity and stochastic processes on the gravitational dynamics have been thoroughly studied [52][53][54][55][56][57], and we will not discuss them further here.

### E. Dynamics of a Uniform Cylindrical Charge Distribution in a Magnetic Field

We now consider the case of the application of an axial constant uniform magnetic field to a cylindrical charge distribution, which directly relates to our previous experimental report on cyclotron dynamics of photoemitted electron gases [28]. It is another example for which the stochastic force must be taken into account. For a uniform cylindrical charge distribution with the density

$$\rho(r,t) = -\frac{n_z e}{\pi R^2(t)} H\left(1 - \frac{r}{R(t)}\right), \tag{51}$$

where $R(t)$ is the radius of the charge distribution, $n_z = -\frac{2\pi}{e} P(\infty, t)$ is the number of electrons per unit length along the z-axis, and $H(\cdot)$ is the Heaviside function in Eq. (24), Eq. (5) simplifies to

$$\frac{d^2 R}{dt^2} + \omega_L^2 \left(R - \frac{S^2}{R^3}\right) = \frac{n_z e^2}{2\pi m \epsilon_0 R} + \frac{16\eta^2}{R^3}, \tag{52}$$

where $\omega_L = eB_0/2m$ is the Larmor frequency and $S$ is a parameter with units of area and determined by the initial conditions of the charge distribution. Comparing Eq. (52) to previously published results reveals that $\eta$ is the emittance of the electron beam, often shown by $\epsilon$ [8]. Furthermore, our theory even predicts the factor "16" in the emittance term in Eq. (52). In literature, based on some observations, it was proposed to replace $\epsilon$ by $4\epsilon$ which is called the "effective emittance" [8, pp. 322]. However, this factor is explicitly derived in our theory.

There are two terms in Eq. (52) that are inversely proportional to $R^3$: one that is due to the application of the external magnetic field and the angular momentum of the distribution and disappears for $\omega_L = 0$, and the one due to the random motions of electrons. In the literature, it is well known that although these two terms affect $R(t)$ in a similar way, they should be distinguished by their origin [58][59]. Here, the distinction takes place naturally because of the way we integrated the random motions into the dynamics in Section III. Alternatively, Vlasov equation has also been employed to analyze the dynamics of uniform cylindrical charge distribution, which one may find in Section 5.3 of [8], with results identical to what our method has produced.

One application of this theory is to evaluate the emittance of photoemitted electrons. When a femtosecond photoemission process takes place, an oblate charge distribution is created which, as we discussed in Section VI. A, becomes prolate in its early stage of dynamics. If a uniform magnetic field is applied normal to the photoemission surface, this oblate-to-prolate shape transition will be faster since the magnetic field transversely confines and focuses the electron gas periodically. For this case, Eq. (52) can approximately model the dynamics of the charge distribution since the aspect ratio of the distribution is large. The Coulomb explosion term has its largest effect around the creation time and becomes negligible as $n_z$ becomes smaller and smaller. In this case, Eq. (49) can be approximated by

$$\frac{d^2 R}{dt^2} + \omega_L^2 R \approx \frac{16\eta^2 + \omega_L^2 S^2}{R^3} \tag{53}$$

The right-hand-side term in Eq. (53) then prevents the charge distribution from a full collapse on the cylinder axis like the gravitational collapse of an ideal gas discussed previously. Without this term, the radius of the charge distribution shows a pure harmonic oscillation. A solution to Eq. (53) is [60]

$$R(t) = 2\sqrt{2\left(\frac{\sigma_v}{2\omega_L}\right)^2 (1 - \cos 2\omega_L(t - t_0)) + \sigma_s^2}, \quad (54)$$

where $\sigma_v$ and $\sigma_s$ are constant, and, $t_0$ determines the initial time or the phase of oscillations. For $\cos \omega_L(t - t_0)$ far from unity and $\sigma_v \gg \omega_L \sigma_s$, $R(t) \approx 2\frac{\sigma_v}{\omega_L}|\sin \omega_L(t - t_0)|$. For $\cos \omega_L(t - t_0) \to 1$, we find $R(t) \to 2\sigma_s$ and hence $S = 4\sigma_s^2$. Note that $2\omega_L$ is the cyclotron angular frequency. Fig 6 (a) shows the radii of a uniform spheroid charge distribution in a magnetic field over time obtained by an *N*-body simulation. In panel (b), we have fitted function $R(t)$, in Eq. (54), to its transvers radius and extracted $\sigma_v$, $\omega_L$, $\sigma_s$ and $t_0$ with a good accuracy.

Time-of-flight spectroscopy has been used to measure the emittance of photoemitted electrons [61]. Here, we propose another method to evaluate the emittance of electron gases since $R(t)$ in Eq. (54) can be directly measured using ultrafast electron microscopy [28]. There, Eq. (54) is derived by a totally different method. Indeed, the photoemitted electron gas in a uniform magnetic field shows periodic focusing and the size of the focus is affected by its emittance. In the simulation results in Fig 6 (b), we see that the minimum values at which the transvers radius periodically reaches increases gradually. That is because the phase space volume, $\eta^2$, is gradually increasing due to thermalization, which is a known phenomenon [8][62]. The thermalization happens every time that the charge distribution collapses transversely, as shown in Fig. 7. In panels (a), (b) and (c) of that figure, the kinetic energy distributions are sketched at various times. Apparently, the kinetic energy distribution is different from what Fig. 3 (b) shows for unifom spherical charge distribution. Comparing Fig. 7 (d) to Fig. 3 (c), we see that the kinetic energy spread becomes significantly larger in the presence of the magnetic field.

## V. CONCLUSIONS

In this work, we developed an analytical approach to determine the dynamics of charge distributions under their own electric field and a stochastic force due to random motions of electrons, and applied it to ellipsoidal, spherical and cylindrical charge distributions. We took advantage of the charge-current continuity equation to derive the current density, which in turn gives the drift velocity and thereby the equation of motion under the Lorentz force. The dynamics derived using this approach are in excellent agreement with results from *N*-body simulations. Since *N*-body simulations come at great computational cost, our analytical approach will be particularly useful to readily study the Coulomb explosion dynamics of, for example, spherical ionized clusters, such as those generated using ultrashort X-ray pulses [63][64], or charged bunches expelled from thin targets [65]. Moreover, to the best of the authors' knowledge, formulating the dynamics of arbitrary (nonuniform) spherical and cylindrical charge distributions, given in the form of integrodifferential equations, is presented for the first time in this work. These formulations include the effect of random motions of electrons through the kinetic theory of emittance that we developed here. We demonstrated some applications of our method, in particular to the gravitational collapse of an ideal gas and the dynamics of a cylindrical charge distribution in a uniform magnetic field. Based on our previous work [28], we proposed an empirical way to measure the emittance of a photoemitted electron gas. Since the method uses the charge-current continuity equation, it is possible to include external current sources such as photoemission. Future efforts will be geared towards reformulating the method for relativistic electrons which is of great interest in the fields of accelerator and plasma physics.

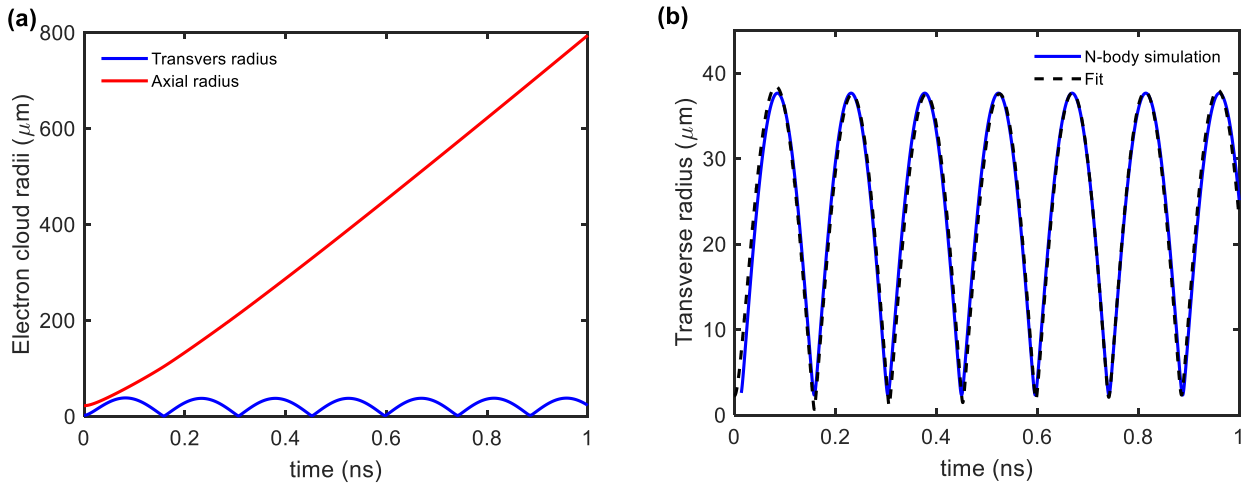

Fig. 6 (a) Transverse and axial radii of a uniform spheroid charge distribution, consisting of 10,000 electrons as a function of time, with the initial values of $\sqrt{5}$ µm and $10\sqrt{5}$ µm, respectively, within an axial uniform magnetic field of B = 0.25 T, determined by an *N*-body simulation. The dynamics starts from rest. (b) The transverse radius of the charge distribution with the function in Eq. (54) fitted to it. The fit parameters are $\omega_L = 21.53$ GHz, $\sigma_s = 0.9204 \mu m$, $\sigma_v = 3.627 \times 10^5$ m/s and $t_0 = 742$ ps. The Larmor frequency is $\omega_L = eB/2m = 21.95$ GHz. The slight detune from this frequency is due to space charge effect. The average of the electrons' velocities spread over all time, obtained from the simulation, is equal $3.681 \times 10^5$ m/s which is close to the fit parameter. The minima of transverse radius of the distribution increase gradually, which indicates that the phase space volume is increasing due to the thermalization.

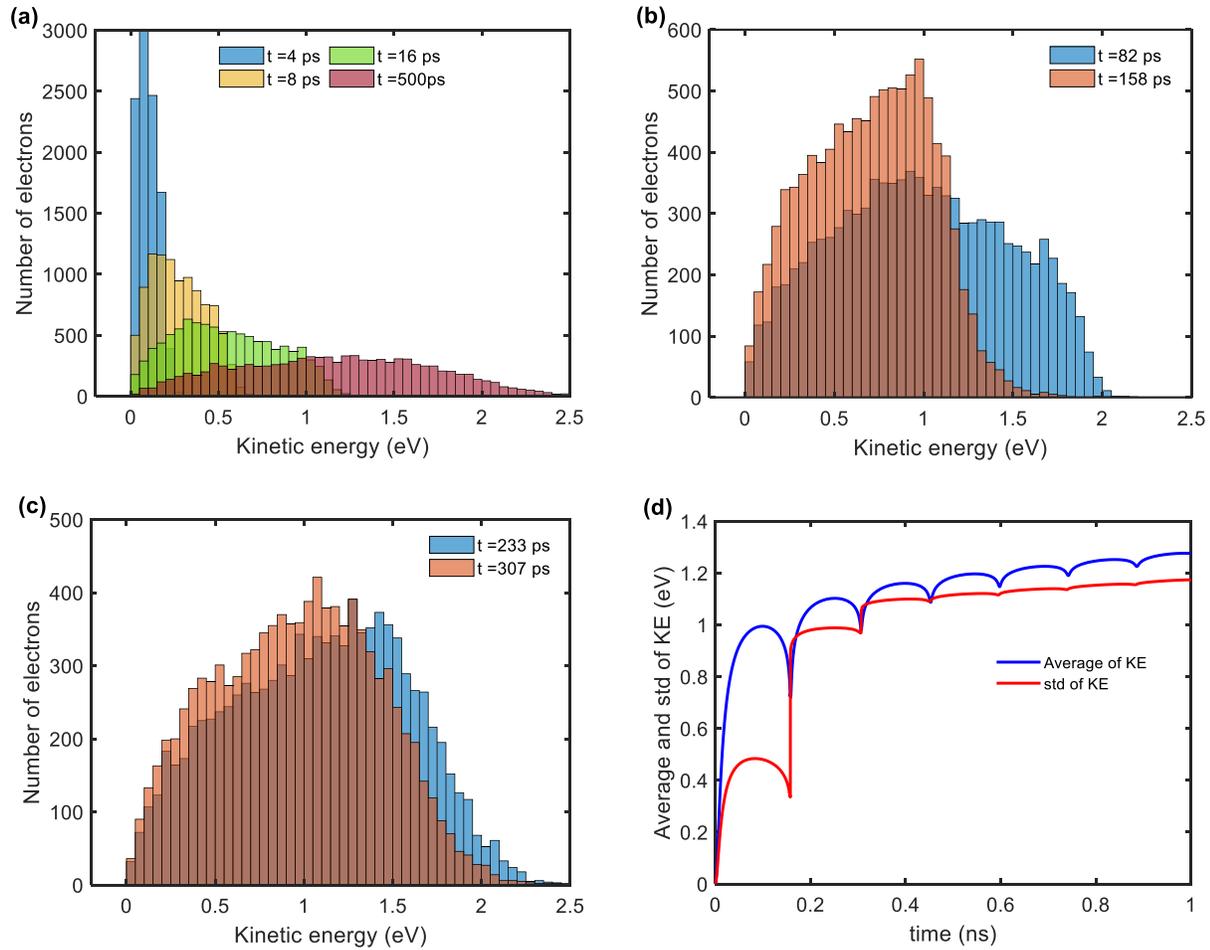

Fig. 7 (a) Number of electrons per kinetic energy at three early and one later time (b) Number of electrons per kinetic energy at the first maximum of the charge distribution radius and the minimum next to it. The kinetic energy distribution gets narrower as the charge density collapses, at which a thermalization happens due to electron-electron interactions. (c) Number of electrons per kinetic energy at the second maximum and the minimum next to it. There is less change in the kinetic energy distribution at those points. The bin width for all histograms is 50 meV. (d) The average and std of the cloud kinetic energy vs time determined by the simulation. There are local minima happening at the minima of transverse radius of the electron cloud.

## ACKNOWLEDGEMENTS

We thank Brandon S. Zerbe, and Phillip M. Duxbury for fruitful discussions. This research was supported by a Packard Fellowship in Science and Engineering from the David and Lucile Packard Foundation.

# APPENDIX A: DERIVATION OF THE NON-LINEAR INTEGRODIFFERENTIAL EQUATION FOR NON-UNIFORM SPHERICAL CHARGE DISTRIBUTIONS

For the general case of a spherically symmetric, nonuniform distribution $\rho(r,t)$, we use Gauss's law to find the current density in terms of the charge density

$$\boldsymbol{J}(r,t) = -\frac{\hat{\boldsymbol{r}}}{r^2}\frac{\partial}{\partial t}\int_0^r \rho(r',t)r'^2 dr'. \tag{A1}$$

Here, $r$ is the radial coordinate variable in the spherical coordinate system. The corresponding drift velocity is

$$\boldsymbol{v} = \frac{\boldsymbol{J}(r,t)}{\rho(r,t)} = -\frac{\frac{\partial}{\partial t}\int_0^r r'^2 \rho(r',t)dr'}{r^2\rho(r,t)}\hat{\boldsymbol{r}}. \tag{A2}$$

Now, we find the material derivative of the drift velocity by calculating terms $\partial\boldsymbol{v}/\partial t$ and $\boldsymbol{v}\cdot\nabla\boldsymbol{v}$. We have

$$\frac{\partial \boldsymbol{v}}{\partial t} = -\hat{\boldsymbol{r}}\frac{\rho(r,t)\frac{\partial^2}{\partial t^2}\int_0^r r'^2\rho(r',t)dr' - \frac{\partial}{\partial t}\rho(r,t)\frac{\partial}{\partial t}\int_0^r r'^2\rho(r',t)dr'}{r^2\rho^2(r,t)} \tag{A3}$$

and

$$\boldsymbol{v}\cdot\nabla\boldsymbol{v} = \hat{\boldsymbol{r}} v\frac{\partial v}{\partial r} = \hat{\boldsymbol{r}}\frac{\frac{\partial}{\partial t}\int_0^r r'^2\rho(r',t)dr'}{r^2\rho(r,t)}\frac{\partial}{\partial r}\left(\frac{\frac{\partial}{\partial t}\int_0^r r'^2\rho(r',t)dr'}{r^2\rho(r,t)}\right). \tag{A4}$$

Evaluation of the spatial derivative on the right-hand-side of Eq. (A4) gives

$$\boldsymbol{v}\cdot\nabla\boldsymbol{v} = \hat{\boldsymbol{r}}\frac{\frac{\partial}{\partial t}\int_0^r r'^2\rho(r',t)dr'}{r^2\rho(r,t)}\left\{\frac{r^2\rho(r,t)\frac{\partial}{\partial t}r^2\rho(r,t) - \left(2r\rho(r,t) + r^2\frac{\partial}{\partial r}\rho(r,t)\right)\frac{\partial}{\partial t}\int_0^r r'^2\rho(r',t)dr'}{r^4\rho^2(r,t)}\right\}. \tag{A5}$$

Adding Eq. (A3) and (A5) will give the material derivative of the drift velocity as

$$\frac{d\boldsymbol{v}}{dt} = \frac{\partial \boldsymbol{v}}{\partial t} + \boldsymbol{v}\cdot\nabla\boldsymbol{v} = -\hat{\boldsymbol{r}}\frac{\rho(r,t)\frac{\partial^2}{\partial t^2}\int_0^r r'^2\rho(r',t)dr' - 2\frac{\partial}{\partial t}\rho(r,t)\frac{\partial}{\partial t}\int_0^r r'^2\rho(r',t)dr'}{r^2\rho^2(r,t)}$$

$$-\hat{\boldsymbol{r}}\left[\frac{2}{r^5\rho^2(r,t)} + \frac{\frac{\partial}{\partial r}\rho(r,t)}{r^4\rho^3(r,t)}\right]\left(\frac{\partial}{\partial t}\int_0^r r'^2\rho(r',t)dr'\right)^2. \tag{A6}$$

By use of Gauss's law, we find the electric field of the charge distribution as

$$\boldsymbol{E}(r,t) = \frac{1}{\epsilon_0}\frac{\hat{\boldsymbol{r}}}{r^2}\int_0^r \rho(r',t)r'^2 dr'. \tag{A7}$$

The non-relativistic equation of motion

$$\frac{d\boldsymbol{v}}{dt} = -\frac{e}{m}\boldsymbol{E}(r,t) \tag{A8}$$

after some algebraic simplifications gives

$$\left[-\rho(r,t)\frac{\partial^2}{\partial t^2} + 2\frac{\partial \rho(r,t)}{\partial t}\frac{\partial}{\partial t} + \frac{e}{m\epsilon_0}\rho^2(r,t)\right]\int_0^r r'^2 \rho(r',t)dr' = \left[\frac{2}{r^3} + \frac{1}{r^2\rho(r,t)}\frac{\partial \rho(r,t)}{\partial r}\right]\left(\frac{\partial}{\partial t}\int_0^r r'^2 \rho(r',t)dr'\right)^2, \quad (A9)$$

which is a non-linear integrodifferential equation containing only $\rho(r,t)$. The equation, though difficult to solve, has the advantage of describing the charge density as a function of both $r$ and $t$, and hence can fully determine the dynamics of the distribution. Generalization to the relativistic expansion is straightforward but beyond the scope of this paper.

# APPENDIX B: COULOMB EXPLOSION OF A CYLINDRICAL CHARGE DISTRIBUTION IN AN AXIAL UNIFORM MAGNETIC FIELD

For the general case of a cylindrically symmetric charge distribution $\rho(r,t)$ in the cylindrical coordinate system $(r,\phi,z)$, we use Gauss's law to find the lateral current density in terms of the charge density

$$\boldsymbol{J}(r,t) = -\frac{\hat{r}}{r}\frac{\partial}{\partial t}P(r,t) + v_\phi(r,t)\rho(r,t)\hat{\phi}, \quad (B1)$$

where

$$P(r,t) = \int_0^r \rho(r',t)r'dr' \quad (B2)$$

and $v_\phi(r,t)$ the rotational velocity of the charge distribution around the magnetic field axis which is determined by the dynamics of the system. The corresponding lateral drift velocity is

$$\boldsymbol{v} = \frac{\boldsymbol{J}}{\rho} = -\frac{1}{r\rho}\frac{\partial P}{\partial t}\hat{r} + v_\phi\hat{\phi}. \quad (B3)$$

Now, we find the material derivative of the drift velocity given by calculating terms $\partial\boldsymbol{v}/\partial t$ and $\boldsymbol{v}\cdot\nabla\boldsymbol{v}$. We have

$$\frac{\partial\boldsymbol{v}}{\partial t} = -\left(\frac{\rho}{r\rho^2(r,t)}\frac{\partial^2 P}{\partial t^2} - \frac{1}{r\rho^2(r,t)}\frac{\partial}{\partial t}\rho\frac{\partial P}{\partial t}\right)\hat{r} + \frac{\partial v_\phi}{\partial t}\hat{\phi}, \quad (B4)$$

and

$$\boldsymbol{v}\cdot\nabla\boldsymbol{v} = \left(\frac{1}{r\rho}\frac{\partial P}{\partial t}\frac{\partial}{\partial r}\left(\frac{1}{r\rho}\frac{\partial P}{\partial t}\right) - \frac{v_\phi^2}{r}\right)\hat{r} + \left(\frac{1}{r\rho}\frac{\partial P}{\partial t}\frac{\partial v_\phi}{\partial r} + \frac{v_\phi}{r^2\rho}\frac{\partial P}{\partial t}\right)\hat{\phi}. \quad (B5)$$

The non-relativistic equation of motion is

$$\frac{d\boldsymbol{v}}{dt} = -\frac{e}{m}\boldsymbol{E} - \frac{e}{m}\boldsymbol{v}\times\boldsymbol{B} = -\frac{e}{m}\boldsymbol{E} - \frac{e}{m}\left(\frac{1}{r\rho}\frac{\partial P}{\partial t}\hat{r} - v_\phi\hat{\phi}\right)\times\boldsymbol{B}. \quad (B6)$$

We first determine the $\hat{\phi}$ component of the equation of motion for the magnetic field $\boldsymbol{B} = B_0 \hat{z}$. By use of Eq. ($B4$) and ($B5$) in Eq. ($B6$), we have

$$\frac{\partial v_\phi}{\partial t} + \frac{1}{r\rho}\frac{\partial P}{\partial t}\frac{\partial v_\phi}{\partial r} + \frac{v_\phi}{r^2\rho}\frac{\partial P}{\partial t} = \frac{e}{m}\frac{1}{r\rho}\frac{\partial P}{\partial t}B_0, \tag{B7}$$

which simplifies to

$$r\rho\left(\frac{\partial P}{\partial t}\right)^{-1}\frac{\partial v_\phi}{\partial t} + \frac{\partial v_\phi}{\partial r} + \frac{v_\phi}{r} = \frac{eB_0}{m}. \tag{B8}$$

The electric field is determined by Gauss's law

$$E_r(r,t) = \frac{1}{\epsilon_0}\frac{1}{r}P \tag{B9}$$

from which the radial component of the equation of motion becomes

$$\left[-\frac{\partial^2}{\partial t^2} + \frac{2}{\rho}\frac{\partial \rho}{\partial t}\frac{\partial}{\partial t} + \frac{e\rho}{m\epsilon_0}\right]P = \left(v_\phi + \frac{eB_0}{m}r\right)v_\phi\rho + \frac{1}{r^2\rho^2}\frac{\partial(r\rho)}{\partial r}\left(\frac{\partial P}{\partial t}\right)^2. \tag{B10}$$

Eq. ($B10$), ($B8$) and ($B2$) must be solved simultaneously to determine the dynamics of a cylindrical charge distribution in a uniform magnetic field along its axis. Eq. ($B8$) shows that, in general, the rotational dynamics of a cylindrical charge distribution is significantly affected by the self-Coulomb field.

## APPENDIX C: N-BODY SIMULATIONS

In the $N$-body simulations, we calculate the Coulomb force on each electron inside an electron cloud due to other electrons. It is assumed that the electrons move much slower than the speed of light (they are non-relativistic) and hence it is not necessary to use retarded fields. Also, electrodynamic radiation is negligible. The increment of time is linear, with $\Delta t$ steps, counted by $n$ as

$$t = (n-1)\Delta t; \quad n = 1, 2, 3, \dots. \tag{C1}$$

At each time instant, the force on the $i$th electron is

$$\boldsymbol{F}_i^{(n)} = \frac{e^2}{4\pi\epsilon_0}\sum_{j\neq i}^{N}\frac{\boldsymbol{r}_i^{(n)} - \boldsymbol{r}_j^{(n)}}{\left|\boldsymbol{r}_i^{(n)} - \boldsymbol{r}_j^{(n)}\right|^3} - e\boldsymbol{v}_i^{(n)} \times \boldsymbol{B}, \tag{C2}$$

where $\boldsymbol{r}_i^{(n)}$ and $\boldsymbol{v}_i^{(n)}$ are the position and velocity of the $i$th electron at the $n$th time step, respectively and $\boldsymbol{B}$ is an external magnetic field. At the initial time when $n = 1$, the position and the velocity of the electrons can have any distribution, which is generated by a random process. The equation of motion is

$$\frac{d\boldsymbol{p}_i^{(n)}}{dt} = \boldsymbol{F}_i^{(n)}, \tag{C3}$$

where $\boldsymbol{p}_i^{(n)}$ is the momentum of the *i*th electron at the *n*th time step. The velocity and position of the *i*th electron at the next time step will be

$$\boldsymbol{v}_i^{(n+1)} = \boldsymbol{v}_i^{(n)} + \frac{\boldsymbol{F}_i^{(n)}}{m}\Delta t \quad \text{and} \quad \boldsymbol{r}_i^{(n+1)} = \boldsymbol{r}_i^{(n)} + \boldsymbol{v}_i^{(n)}\Delta t, \tag{C4}$$

respectively, which will be used to calculate $\boldsymbol{F}_i^{(n+1)}$. The iteration continuous until the end of the simulation time. A step of 80 fs is selected as the smallest value, where any further reduction does not increase the accuracy of the simulation considerably.

## APPENDIX D: SPHERICAL CHARGE DISTRIBUTIONS WITH LINEAR-IN-SPACE CURRENT DENSITIES

Suppose we have a spherically symmetric charge distribution of the following form

$$\rho(r,t) = -Ne\frac{f(r,\sigma_r(t))}{\frac{4}{3}\pi\sigma_r^3(t)}, \tag{D1}$$

where $f(r,\sigma_r(t))$ is a unitless function, the integral of which over whole space is unity. Let us define a current density as

$$\boldsymbol{J} = \frac{\dot{\sigma}_r}{\sigma_r} r \rho(r,t)\hat{\boldsymbol{r}}. \tag{D2}$$

The goal is to use the continuity equation $\nabla \cdot \boldsymbol{J} = -\frac{\partial}{\partial t}\rho(r,t)$ to see under which conditions the definition in Eq. (D2) holds. The time-derivative of the charge density in Eq. (B1) is

$$\frac{\partial}{\partial t}\rho(r,t) = Ne\frac{3\dot{\sigma}_r f(r,\sigma_r(t))}{\frac{4}{3}\pi\sigma_r^4(t)} - \frac{Ne}{\frac{4}{3}\pi\sigma_r^3(t)}\frac{\partial f(r,\sigma_r(t))}{\partial t}. \tag{D3}$$

The divergence of the current density in Eq. (D2) is

$$\nabla \cdot \boldsymbol{J} = \frac{\dot{\sigma}_r}{\sigma_r}\frac{1}{r^2}\frac{\partial}{\partial r}(r^3\rho(r,t)) = \frac{\dot{\sigma}_r}{\sigma_r}\frac{1}{r^2}\left(3r^2\rho(r,t) + r^3\frac{\partial}{\partial r}\rho(r,t)\right)$$

$$= -Ne\frac{\dot{\sigma}_r}{\sigma_r}\left(3\frac{f(r,\sigma_r(t))}{\frac{4}{3}\pi\sigma_r^3(t)} + \frac{r}{\frac{4}{3}\pi\sigma_r^3(t)}\frac{\partial}{\partial r}f(r,\sigma_r(t))\right). \tag{D4}$$

For the continuity equation to hold, we equate Eq. (B3) to the negative of Eq. (D4) and get

$$\frac{\partial f(r,\sigma_r(t))}{\partial t} = -\frac{\dot{\sigma}_r}{\sigma_r}r\frac{\partial}{\partial r}f(r,\sigma_r(t)). \tag{D5}$$

The left-hand-side of Eq. (D5) can be rewritten as

$$\frac{\partial f(r,\sigma_r(t))}{\partial t} = \frac{\partial f(r,\sigma_r(t))}{\partial \sigma_r(t)}\dot{\sigma}_r, \tag{D6}$$

which simplifies Eq. (D5) to

$$\sigma_r \frac{\partial f(r,\sigma_r(t))}{\partial \sigma_r(t)} = -r \frac{\partial}{\partial r} f(r,\sigma_r(t)). \tag{D7}$$

Eq. (D7) demands that

$$f(r,\sigma_r(t)) = f\left(\frac{r}{\sigma_r(t)}\right), \tag{D8}$$

because

$$\frac{\partial}{\partial r} f\left(\frac{r}{\sigma_r(t)}\right) = \frac{df}{d\left(\frac{r}{\sigma_r(t)}\right)} \frac{\partial}{\partial r}\left(\frac{r}{\sigma_r(t)}\right) = \frac{df}{d\left(\frac{r}{\sigma_r(t)}\right)} \frac{1}{\sigma_r(t)} \tag{D9}$$

and

$$\frac{\partial}{\partial \sigma_r(t)} f\left(\frac{r}{\sigma_r(t)}\right) = \frac{df}{d\left(\frac{r}{\sigma_r(t)}\right)} \frac{\partial}{\partial \sigma_r(t)}\left(\frac{r}{\sigma_r(t)}\right) = \frac{df}{d\left(\frac{r}{\sigma_r(t)}\right)} \frac{-r}{\sigma_r^2(t)}. \tag{D10}$$

Therefore, any distribution in the form of

$$\rho(r,t) = \frac{f\left(\frac{r}{\sigma_r(t)}\right)}{\frac{4}{3}\pi\sigma_r^3(t)} \tag{D11}$$

has a corresponding current density in Eq. (D2).